\documentstyle[prl,aps]{revtex}
\title{VAN DER WAALS
EXCLUDED VOLUME MODEL OF MULTICOMPONENT HADRON GAS}

\author{M.I. Gorenstein$^{1,2}$, A.P. Kostyuk$^2$ and  Ya.D. Krivenko$^2$}

\address{
$^1$ Institut f\"ur Theoretische Physik, Universit\"at Frankfurt, Germany\\
$^2$ Bogolyubov Institute for Theoretical Physics,
252143, Kiev, Ukraine
 }

\begin{document}

\maketitle

\begin{abstract}
A generalization of the Van der Waals excluded volume
procedure for the multicomponent hadron gas is proposed.
The derivation is
based on the grand canonical partition function for the system of
particles of several species interacting by hard core potentials. The
obtained formulae for thermodynamical quantities are consistent with
underlying principles of statistical mechanics as well as with
thermodynamical identities.
The model can be applied to the analysis of experimental data
for particle number ratios in relativistic nucleus-nucleus
collisions.
\end{abstract}

\pacs{PACS.\ numbers: 25.75.+r }

\section{Introduction}

Thermal hadron gas (HG) models have recently been used
to fit the data of particle yields in
nucleus-nucleus (A+A) collisions at the AGS and SPS energies
(see e.g. \cite{St:96}).
The ideal HG model becomes  inadequate at the chemical
freeze-out in high-energy A+A collisions:
temperature and baryonic chemical potential obtained from
fitting the particle number ratios at the AGS and SPS energies lead to
artificially large values of total particle number and
energy densities (see e.g. \cite{Yen:98}). This is hardly consistent
with a picture of a gas of point-like noninteracting hadrons.

The Van der Waals (VdW) excluded volume procedure appeared to be
effective in taking into account the hadron repulsion at short distances.
It suppresses undesirable large values of
particle number densities. Different versions of the VdW HG models
were proposed and applied for fitting experimental data
on particle number ratios in A+A collisions at the AGS and SPS
energies [3-12].
The proper volume of the $i$-th hadron species is expressed in terms
of its hard-core radius $R_i$. An introduction of the phenomenological
parameters $R_i$ changes the particle number ratios in
comparison with ideal HG results.
The VDW model formulation, however, has not been
properly defined in the case when the $R_i$s are not equal
to each other.

The aim of the present paper is
to propose a generalization of the VdW excluded volume
procedure for the HG gas of several particle species
with different particle radii.  The
derivation is based on the grand canonical partition function for the
system of particles of several species interacting by hard-core
potentials. The obtained formulae are therefore consistent with
the underlying principles of statistical mechanics as well as with
thermodynamical identities.  The pressure, particle densities and other
thermodynamical quantities as functions of temperature and chemical
potentials are defined by a set of coupled transcendental
equations.

\section{One component VDW Gas}

The canonical partition function (CPF) for the one-component
classical (Boltzmann) gas can be written as
\begin{eqnarray}\label{cpartp1}
Z(V,T,N) ~=~ \frac{1}{N!}
\int ~ \prod_{i=1}^{N}\frac{d^3 p_i d^3 r_i }
{(2\pi)^3}
\exp\left(- \frac{\sqrt{ p_i^2 +m_2}}{T} - \frac{U}{T} \right)
\end{eqnarray}
where $V$ and $T$ are the system volume and temperature,
$m$ and $N$  are the mass and number of particles, respectively.
The function
$U$ in Eq.~(\ref{cpartp1}) is assumed to be equal
to the sum of pair potentials:
\begin{equation}\label{potential1}
U ~=~ \sum_{1\leq i < j \leq N}
u (|\vec{r}_{i} - \vec{r}_{j}|)~.
\end{equation}
After integration over the particle momenta,
Eq.~(\ref{cpartp1}) is reduced to
\begin{equation}\label{cpart2}
Z(V,T,N)~ =~ \frac{1}{N!}~
\left[ \phi (T;m) \right ]^{N}~
\int ~ \prod_{i=1}^{N}  d^3 r_i~
\exp\left(-\frac{U}{T}\right)~.
\label{cpart1}
\end{equation}
Here we use the notation
\begin{equation}\label{phi}
\phi(T;m)~ =~ \frac{1}{2 \pi^2} \int_0^{\infty}p^2 dp~
\exp \left( - \frac{\sqrt{p^{2}+m^{2}}}{T}  \right)~
= ~\frac{m^{2} T}{2 \pi^{2}}~ K_{2}\left( \frac{m}{T} \right),
\end{equation}
where $K_{2}$ is the modified Bessel function.
The asymptotic behavior of $\phi(T;m)$ in the non-relativistic
limit, $m>>T$, has the form
\begin{equation}
\phi(T;m)~\cong~
\left( \frac{m T}{2 \pi} \right)^{3/2}~ \exp(-m/T)~.
\end{equation}

By means of the Mayer functions
\begin{equation}\label{fij1}
f_{ij}~\equiv ~\exp\left(
-\frac{u(|\vec{r}_{i}-\vec{r}_{j}|)}{T}
\right)~ -~ 1~,
\end{equation}
one can rewrite the integrand of Eq.~(\ref{cpart1}) in the following form
\begin{equation}\label{rinteg1}
\exp\left(- \frac{U}{T}\right)~ =~
\prod_{i=1}^{N-1}  \prod_{j=i+1}^{N} [ 1+f_{ij} ]~
\cong~
\prod_{i=1}^{N-1} [ 1+ \sum_{j=i+1}^{N} f_{ij} ]
\end{equation}
The approximate equality in the last expressions is based on the
assumption that the gas is rarefied and, therefore,
the terms containing products $f_{ij} f_{im}$ can be dropped.

Let us introduce the notations
\begin{equation}\label{b}
\int d^{3} r_{i} f_{ij}~ =~ - ~2 b(T,\vec{r}_{j})~.
\end{equation}
Now we shall use the rigid ball model, i.e. we assume the
hard-core interaction between the particles:
\begin{equation}
u(|\vec{r}_{i}-\vec{r}_{j}|)  =
\left\{
\begin{array}{lcl}
\infty & \mbox{ if } &
|\vec{r}_{i}-\vec{r}_{j}| \le 2 R \\
   0   & \mbox{ if } &
|\vec{r}_{i}-\vec{r}_{j}| > 2 R ,
\end{array}
\right.
\end{equation}
($R$ is the particle radius). In this
case $b$ defined by Eq.~(\ref{b})
does not depend on temperature. For
$V^{1/3}
\gg R$ the dependence of $b$ on $\vec{r}_{j}$ is negligible
and the integral in Eq.~(\ref{b}) can be easily
calculated:
\begin{equation}\label{b0}
b ~=~ \frac{16}{3} \pi  R^{3}~.
\end{equation}

Substituting Eqs.~(\ref{rinteg1}-\ref{b}) into
Eq.~(\ref{cpart2})
one gets
\begin{eqnarray}\label{cpart3}
Z(V,T,N) ~&=&~ \frac{1}{N!}
\left[ \phi (T;m) \right ]^{N}
\prod_{i=1}^{N} (V - 2 (N - i ) b )
\label{cpartl1} \\
&=&~ \frac{1}{N!}
\left[ \phi (T;m) \right ]^{N}
V^N \exp \left[
\sum_{i=1}^{N} \log \left(1 - 2 (N - i ) \frac{b}{V}
\right) \right]~.
\nonumber
\end{eqnarray}
In a rarefied gas the value
of $2bN$ is much smaller then the total volume $V$.
In this case one can approximate $\log(1-x)\cong -x$
in Eq.~(\ref{cpart3}).
This yields
\begin{eqnarray}
Z(V,T,N)
~\cong~ \frac{1}{N!}~
\left[ \phi (T;m) ~ V\right ]^{N}
~\exp \left(
- \sum_{i=1}^{N} 2 (N - i ) \frac{b}{V}~
\right)~.
\end{eqnarray}
Performing the summation in the exponential
and using approximate relation $\exp(-Nb/V)\cong 1-Nb/V$
one gets CPF
of one component rigid ball gas:
\begin{eqnarray}\label{cpfa1}
Z(V,T,N)
~&\cong&~ \frac{1}{N!}
\left[ \phi (T;m)~V \right ]^{N}
~\exp \left(
- N (N - 1) \frac{b}{V}
\right)
\nonumber \\
&\cong&~
\frac{1}{N!}~
\left[ \phi (T;m) \right ]^{N}~ (V - N b )^{N}~.
\label{cparte1}
\end{eqnarray}
Substituting the expression (\ref{cpfa1}) into the formula
for pressure
one finds the well known VdW equation for rigid ball
gas:
\begin{equation}\label{steq1}
p(V,T,N)~ \equiv~
T~ \frac{\partial \log Z(V,T,N)}{\partial V}
~=~\frac{N T}{V - b N  }~.
\end{equation}
Note that $b$ (\ref{b0}) is equal to $4v$, where $v=4\pi R^3/3$
is the particle volume.
The VdW equation of state (\ref{steq1}) is obtained
from the statistical mechanics of rigid balls within
a rarefied gas approximation,
$vN<<V$. For a dense gas
the VDW equation (\ref{steq1}) should be considered as a phenomenological
extrapolation.

The grand canonical partition function (GCPF) is expressed through the CPF
in the following way:
\begin{equation}
{\cal Z}(V,T,\mu)~ =~
\sum_{N=0}^{\infty}
\exp \left( \frac{\mu N}{T} \right)
Z(V,T,N)~,
\label{gcpfe1}
\end{equation}
where $\mu$ is the chemical potential.
In the case of rigid ball gas, the upper limit of the sum is
in fact not infinite. The CPF
(\ref{cpart1}) becomes equal to zero if $N$ exceeds
$N_{0}\sim V/v$.

Let $\exp \left( \frac{\mu N^{*}}{T} \right) Z(V,T,N^{*})$
is the maximal
term in the sum (\ref{gcpfe1}). The rest of terms are also
positive, hence the following inequalities are satisfied
\begin{equation}
\exp\left(\frac{\mu N^*}{T}\right)~Z(V,T,N^{*}) ~<~ {\cal Z}(V,T,\mu)~ <~
N_{0}~\times~\exp\left(\frac{\mu N^*}{T}\right)~ Z(V,T,N^{*})~.
\label{ineqz1}
\end{equation}
The pressure can be expressed via GCPF by the formula
\begin{equation}\label{pmu1}
p(T,\mu)~ =~
T~ \lim_{V \rightarrow \infty} \frac{ \log {\cal Z} (V,T,\mu)}{V}~.
\end{equation}
Taking Eq.~(\ref{ineqz1}) into account one gets
\begin{eqnarray}
p(T,\mu)~&\ge &~\lim_{V \rightarrow \infty}
~\frac{\mu N^*+T \log
Z(V,T,N^{*})}{V}~, \nonumber \\
p(T,\mu) ~&\le & ~
\lim_{V \rightarrow \infty}
\frac{ \mu N^*+T\log
Z(V,T,N^{*})}{V}
~+~T~\lim_{V \rightarrow \infty}\frac{\log N_{0}}{V}~.\nonumber
\end{eqnarray}
Since $V^{-1}\log N_{0} \rightarrow 0$ in the thermodynamical limit,
$V\rightarrow \infty$, the pressure $p(T,\mu)$ (\ref{pmu1}) is defined by
the largest
term, with $N=N^*$, of the GCPF (\ref{gcpfe1}).
$N^*$ is also the average number of particles
in the grand canonical formulation.
Using the VdW  approximation
(\ref{cparte1}) for the CPF one finds
\begin{equation}\label{pnsvdw1}
p(T,\mu)~ =~ T~
\lim_{V \rightarrow \infty} \frac{1}{V}~
\log \left[
\frac{A^{N^{*}}  (V - N^{*} b )^{N^{*}} }{N^{*}!}
\right],
\end{equation}
where $A=\exp( \mu/T ) \phi(T;m)$, and $N^*=N^*(V,T,\mu)$
corresponds to the maximum of the expression in the square brackets.

Let us show that Eq.~(\ref{pnsvdw1}) leads to the result
\begin{equation}\label{pxi1}
p(T,\mu)~ =~ T \xi~,
\end{equation}
where $\xi$ is defined by
the transcendental equation
\begin{equation} \label{eqxi}
 \xi~ =~ A ~\exp(- b \xi)~.
\end{equation}
Using the asymptotic representation
for the $\Gamma$-function logarithm at $N \rightarrow \infty$
\begin{equation}
\log \Gamma (N + 1) \cong N ( \log N - 1)
\label{Gamas}
\end{equation}
it is easy to check that the value of $N^{*}$ satisfying the
maximum condition of logarithm argument in Eq.~(\ref{pnsvdw1})
is given by the formula
\begin{equation}\label{ns1}
N^{*}~ \cong~ V n~,
\end{equation}
where $n=n(T,\mu)$ is related to $\xi$ via equation
\begin{equation}\label{eqn}
n~ =~ \frac{\xi}{1 + b \xi}~.
\end{equation}
Substitution of Eq.~(\ref{ns1}) into Eq.~(\ref{pnsvdw1}) with account for
Eqs.(\ref{eqxi},\ref{eqn})  yields the
formula (\ref{pxi1}).
The quantity $n=n(T,\mu)$
is a particle
number density in the grand canonical formulation. One can readily check
that the definition of the particle number density,
$n=\partial p(T,\mu)/\partial \mu$~,
leads to Eq.~(\ref{eqn}), provided that
Eqs.~(\ref{pxi1}-\ref{Gamas})
are taken into account.

For the point like particles, $R=0$ and  $b=0$,
Eq.~(\ref{pxi1}) is reduced to the ideal gas
result:
\begin{equation}\label{ideal}
p^{id}(T,\mu)~=~T~\exp(\mu/T)~\phi(T;m)~=~T~n^{id}(T,\mu)~.
\end{equation}
Eq.~(\ref{pxi1}) can be therefore written in the form \cite{ris91}:
\begin{equation}\label{nonideal}
p(T,\mu)~=~
p^{id}\left(T,\mu-\frac{bp(T,\mu)}{T}\right)~.
\end{equation}
It can be also presented in the form of Eq.~(\ref{steq1})
with $N=N^*(V,T,\mu)$, which demonstrates explicitly
the equivalence between canonical and grand canonical
formulations at $V\rightarrow \infty$.

\section{Two Component VDW Gas}

In the case of two particle species CPF has the following form:
\begin{eqnarray}\label{cpartp}
Z(V,T,N_{1},N_{2})& =& \frac{1}{N_{1}! N_{2}!}
\int ~ \prod_{i=1}^{N_1}\frac{d^3 p_i^{(1)} d^3 r_i^{(1)}}
{(2\pi)^3}
\exp\left(- \frac{\sqrt{m_1^2 + (p_i^{(1)})^2}}{T}\right)\\
&  \times & \prod_{k=1}^{N_2}\frac{d^3 p_k^{(2)} d^3 r_k^{(2)}}{(2\pi)^3}
\exp\left(- \frac{\sqrt{m_2^2 + (p_k^{(2)})^2}}{T}\right) ~
\exp\left(-\frac{U^{(1,2)}}{T}\right) \nonumber ~,
\end{eqnarray}
where
$m_1$, $N_{1}$
($m_2$, $N_{2}$) are the mass and number of particles of the 1-st
(2-nd) species,
\begin{eqnarray}\label{potential}
U^{(1,2)}~& =&~ \sum_{1\leq i < j \leq N_{1}}
u_{11} (|\vec{r}_{i}^{(1)}-\vec{r}_{j}^{(1)}|)
~+~ \sum_{1\leq k < l \leq N_{2}}
u_{22} (|\vec{r}_{k}^{(2)}-\vec{r}_{l}^{(2)}|)
\\
&+&~  \sum_{i=1}^{N_{1}} \sum_{k=1}^{N_{2}}
u_{12} (|\vec{r}_{i}^{(1)}-\vec{r}_{k}^{(2)}|)~.\nonumber
\end{eqnarray}

It should be mentioned that in the case of {\it two} particle species
$U^{(1,2)}$ contains {\it three} types of two-particle potentials.
While the potentials $u_{11}$ and $u_{22}$ describe interactions
between particles of the same species and can be handled
similarly to the potential $u$ of one-component case, the
potential $u_{12}$ describing interactions between particles of
different species requires an special treatment and prevents
a straightforward generalization of the one-component VDW
equation to the two-component gas.

The integration of (\ref{cpartp}) over the particle momenta
gives the
following expression for the CPF
\begin{equation}
Z(V,T,N_{1},N_{2}) = \frac{\left[ \phi (T;m_{1}) \right ]^{N_{1}}
\left[ \phi (T;m_{2}) \right ]^{N_{2}}}{N_{1}!
N_{2}!}
~\int ~ \prod_{i=1}^{N_1}  d^3 r_i^{(1)}
\prod_{k=1}^{N_2} d^3 r_k^{(2)}
\exp\left(-\frac{U^{(1,2)}}{T}\right)~.
\label{cpart}
\end{equation}
The next step, however, involves the Mayer functions of
{\it three} types ($p,q$=1,1;2,2;1,2):
\begin{equation}\label{fij}
f^{(pq)}_{ik}~\equiv~\exp\left(
-\frac{u_{pq}(|\vec{r}_{i}^{(p)}-\vec{r}_{k}^{(q)}|)}{T}
\right)~-~1~,
~~~~p,q=1,2~
\end{equation}
(note that $f^{12}_{il}=f^{21}_{li}$).
The integrand of (\ref{cpart}) can be rewritten as
\begin{eqnarray}\label{rinteg}
\exp\left(- \frac{U}{T}\right) &=&
\prod_{k=1}^{N_{w}-1}  \prod_{l=k+1}^{N_{w}} [ 1+f^{(ww)}_{kl} ]~
\prod_{i=1}^{N_{v}} \prod_{j=i+1}^{N_{v}} [1+f^{(vv)}_{ij}]~
\prod_{m=1}^{N_{w}} [1+f^{(vw)}_{im}] \\
& \cong &
\prod_{k=1}^{N_{w}-1} [ 1+ \sum_{l=k+1}^{N_{w}} f^{(ww)}_{kl} ]~
\prod_{i=1}^{N_{v}} [1+ \sum_{j=i+1}^{N_{v}} f^{(vv)}_{ij}
+ \sum_{m=1}^{N_{w}} f^{(vw)}_{im}] ,
\nonumber
\end{eqnarray}
where $v,w~=~1,2$ and $N_{v} \ge N_{w}$.

For each type of Mayer function we introduce the notation
\begin{equation}\label{bpq}
\int d^{3} r^{(p)}_{i} f^{(pq)}_{ij} = - 2 b_{pq}(T,\vec{r}^{q}_{j}).
\end{equation}
For the rigid ball model at $V^{1/3} \gg max(R_1,R_2)$
($R_p$ is the radius of a particle of $p$-th species, $p=1,2$) it
yields
\begin{equation}\label{bpq0}
b_{pq} = \frac{2}{3} \pi (R_p+R_q)^{3}~.
\end{equation}

Substituting (\ref{rinteg}) and (\ref{bpq}) into (\ref{cpart})
one gets
\begin{eqnarray}\label{zn1n2}
&& Z(V,T,N_{1},N_{2}) = \frac{1}{N_{1}! N_{2}!}
\left[ \phi (T;m_{1}) \right ]^{N_{1}}
\left[ \phi (T;m_{2}) \right ]^{N_{2}}
\prod_{k=1}^{N_{w}} (V - 2 (N_{w} - k ) b_{ww} )  \\
& & \times
\prod_{i=1}^{N_{v}} (V - 2 (N_{v} - i) b_{vv} - 2 N_{w} b_{vw} )
\label{cpartl}
= \frac{1}{N_{1}! N_{2}!}
\left[ \phi (T;m_{1}) \right ]^{N_{1}}
\left[ \phi (T;m_{2}) \right ]^{N_{2}}
V^{N_{1} + N_{2}}\nonumber \\
& & \times \exp \left[
\sum_{k=1}^{N_{w}} \log \left(1 - 2 (N_{w} - k ) \frac{b_{ww}}{V}
\right)
 +
\sum_{i=1}^{N_{v}} \log \left(1 - 2 (N_{v} - i) \frac{b_{vv}}{V}
- 2 N_{w} \frac{b_{vw}}{V} \right)
\right]~.
\nonumber
\end{eqnarray}
We again assume that the gas is rarefied and the total proper volume of
all particles is much smaller then the total volume
of the system.
Then it follows from Eq.~(\ref{zn1n2}):
\begin{eqnarray}
Z(V,T,N_{1},N_{2})
&\cong & \frac{1}{N_{1}! N_{2}!}
\left[ \phi (T;m_{1}) \right ]^{N_{1}}
\left[ \phi (T;m_{2}) \right ]^{N_{2}}
V^{N_{1} + N_{2}} \\
& & \times
\exp \left[
- \sum_{k=1}^{N_{w}} 2 (N_{w} - k ) \frac{b_{ww}}{V}
- \sum_{i=1}^{N_{v}} \left( 2 (N_{v} - i) \frac{b_{vv}}{V}
+ 2 N_{w} \frac{b_{wv}}{V} \right)
\right]~. \nonumber
\end{eqnarray}
Performing the summation in the exponential
\begin{eqnarray}
Z(V,T,N_{1},N_{2})
&\cong& \frac{1}{N_{1}! N_{2}!}
\left[ \phi (T;m_{1}) \right ]^{N_{1}}
\left[ \phi (T;m_{2}) \right ]^{N_{2}}
V^{N_{1} + N_{2}}
\nonumber \\
& & \times
\exp \left(
- N_{2} (N_{2} - 1) \frac{b_{22}}{V}
- N_{1} (N_{1} - 1) \frac{b_{11}}{V}
- 2 N_{1} N_{2} \frac{b_{12}}{V}
\right) \label{cparte}
\end{eqnarray}
and introducing quantities $\tilde{b}_{12}$ and $\tilde{b}_{21}$
constrained by the condition
\begin{equation}\label{constr1}
\tilde{b}_{12} + \tilde{b}_{21} =2 b_{12}~,
\end{equation}
we can rewrite the CPF in the form
\begin{eqnarray}
Z(V,T,N_{1},N_{2})
&\cong&
\frac{1}{N_{1}! N_{2}!}
\left[ \phi (T;m_{1}) \right ]^{N_{1}}
\left[ \phi (T;m_{2}) \right ]^{N_{2}}
V^{N_{1} + N_{2}}
\left[ \exp \left(
- N_{1} \frac{b_{11}}{V}
- N_{2} \frac{\tilde{b}_{21}}{V}
\right) \right]^{N_{1}} \nonumber \\
& & \times
\left[ \exp \left(
- N_{2} \frac{b_{22}}{V}
- N_{1} \frac{\tilde{b}_{12}}{V}
\right) \right]^{N_{2}}~. \label{cpartet}
\end{eqnarray}
Imposing the additional constraints
$2 N_{v}\tilde{b}_{vw}/V \ll 1$
we obtain the final
expression for the partition function of the two-component
VdW gas
\begin{eqnarray}\label{cpfa}
Z(V,T,N_{1},N_{2})
~&\cong&~
\frac{1}{N_{1}! N_{2}!}
\left[ \phi (T;m_{1}) \right ]^{N_{1}}
\left[ \phi (T;m_{2}) \right ]^{N_{2}} \nonumber \\
&& \times
(V - N_{1} b_{11} - N_{2} \tilde{b}_{21})^{N_{1}}
(V - N_{2} b_{22} - N_{1} \tilde{b}_{12})^{N_{2}}~.
\end{eqnarray}
Substituting expression (\ref{cpfa}) into the formula for the pressure
one obtains
\begin{equation}
p(V,T,N_{1},N_{2})\equiv
T \frac{\partial \log Z(V,T,N_{1},N_{2})}{\partial V}
= \frac{N_{1}T}{V -  b_{11}N_{1} -  \tilde{b}_{21}N_{2} } +
\frac{N_{2}T}{V -  b_{22}N_{2} - \tilde{b}_{12}N_{1} }~.
\label{steq}
\end{equation}
The equation of state (\ref{steq}) is consistent with the virial expansion
up to second
order and can be considered as a generalization of the
VdW equation for the two-component system.
We define $\tilde{b}_{12}$ and $\tilde{b}_{21}$ as the following:
\begin{equation}\label{btilde}
\tilde{b}_{12}~ =~ 2 \frac{b_{11}
b_{12}}{b_{11}+b_{22}}~,~~~
\tilde{b}_{21}~ = ~2 \frac{b_{22} b_{12}}{b_{11}+b_{22}}~.
\end{equation}
Eq.~(\ref{btilde}) satisfies
the constraints (\ref{constr1}) and leads to the correct physical
behavior in the limiting cases when both particle radii are equal to each 
other or when one of the particle
radius is equal to zero.
If two species have equal radii, $R_1=R_2$, equation (\ref{steq})
is reduced to the one-component VdW equation (\ref{steq1})
with $N=N_1+N_2$ and $b=b_{11}=b_{22}=\tilde{b}_{12}=\tilde{b}_{21}$.
Note that in a general case, $R_1\neq R_2$~, the VdW excluded volumes
in Eq.~(\ref{steq}) are
different for different particle species.

The transformation to the grand canonical ensemble is  similar
to that of one-component case. The GCPF has the form
\begin{equation}
{\cal Z}(V,T,\mu_{1},\mu_{2}) ~=~
\sum_{N_{1}=0}^{\infty}
\sum_{N_{2}=0}^{\infty}
\exp \left( \frac{ \mu_{1}N_{1} + \mu_{2}N_{2} } {T}
\right)~
Z(V,T,N_{1},N_{2})~,
\label{gcpfe}
\end{equation}
 where $\mu_{q}$ ($q=1,2$) are the chemical potentials of each particle
species. The total number of terms, $N_{0}$, in double sum (\ref{gcpfe})
is of the order of $V^2$.
Since $V^{-1}\log N_{0}\rightarrow 0$
at $ V \rightarrow \infty$ the pressure can be
expressed via the maximum term
in the sum (\ref{gcpfe}):
\begin{equation}\label{pns}
p(T,\mu_{1},\mu_{2})~ =
 \lim_{V \rightarrow \infty}
\frac{T}{V} \log \left[
\exp \left(\frac{\mu_{1}N_{1}^{*} + \mu_{2}N_{2}^{*}}{T} \right)
Z(V,T,N_{1}^{*},N_{2}^{*})  \right]~.
\end{equation}
In the VDW approximation (\ref{cpfa}) the last  expression
takes the form
\begin{equation}\label{pnsvdw}
p(T,\mu_{1},\mu_{2}) =
 \lim_{V \rightarrow \infty}
\frac{T}{V}~ \log \left[
\frac{A_{1}^{N_{1}^{*}} A_{2}^{N_{2}^{*}}
(V - N_{1}^{*} b_{11} - N_{2}^{*} \tilde{b}_{21} )^{N_{1}^{*}}
(V - N_{2}^{*} b_{22} - N_{1}^{*} \tilde{b}_{12} )^{N_{2}^{*}}
}{N_{1}^{*}! N_{2}^{*}!}
\right]~,
\end{equation}
where $A_{p}=\exp( \mu_{p}/T ) ~\phi(T,m_{p})$~.

Let us show that the pressure (\ref{pnsvdw}) can be calculated by the
formula
\begin{equation}\label{pxi}
p(T,\mu_{1},\mu_{2}) = T(\xi_{1} +\xi_{2})~,
\end{equation}
where the values of $\xi_{q}$ are found from the set of coupled
transcendental equations
\begin{eqnarray}
 \xi_{1} &=& A_{1} \exp(- b_{11} \xi_{1} - \tilde{b}_{12} \xi_{2})~,
\label{eqxi1}\\
 \xi_{2} &=& A_{2} \exp(- b_{22} \xi_{2} - \tilde{b}_{21} \xi_{1})~.
\label{eqxi2}
\end{eqnarray}

Using the asymptotic representation for the $\Gamma$-function
logarithm
it is easy to check that the values of $N_{p}^{*}$ satisfying the
maximum condition of the logarithm argument in Eq.(\ref{pnsvdw})
are given by the formula
\begin{equation}\label{ns}
N_{p}^{*}~\cong~ V n_{p}~,
\end{equation}
where $n_{p}=n_p(T,\mu_1,\mu_2)$ are related to $\xi_{p}$ via 
the equations
\begin{eqnarray}
\xi_{1}~ =~ \frac{n_{1}}{1 - n_{1} b_{11} - n_{2} \tilde{b}_{21}}~,
\label{eqn1}\\
\xi_{2}~ = ~ \frac{n_{2}}{1 - n_{2} b_{22} - n_{1} \tilde{b}_{12}}~.
\label{eqn2}
\end{eqnarray}
The substitution of Eq.~(\ref{ns}) into Eq.~(\ref{pnsvdw})
yields the formula (\ref{pxi}).

The set of linear equations (\ref{eqn1}) and (\ref{eqn2}) can be
solved for $n_{p}$:
\begin{eqnarray}
n_{1}~ &=&~
\frac{\xi_{1}[ 1 + \xi_{2} (b_{22}-\tilde{b}_{21}) ] }
{1 + \xi_{1} b_{11} + \xi_{2} b_{22} +
\xi_{1} \xi_{2} (b_{11} b_{22} - \tilde{b}_{12} \tilde{b}_{21})}
\label{n1}~,\\
n_{2}~ &=&~
\frac{\xi_{2}[1 + \xi_{1}(b_{11}-\tilde{b}_{12})]}
{1 + \xi_{1} b_{11} + \xi_{2} b_{22} +
\xi_{1} \xi_{2} (b_{11} b_{22} - \tilde{b}_{12} \tilde{b}_{21})}~.
\label{n2}
\end{eqnarray}
Similarly to the one component case, the quantities $n_{p}$
are particle number densities.
One can readily check that the definition
$n_{p} = \partial p(T,\mu_{1},\mu_{2})/\partial \mu_{p}$
leads to Eqs.~(\ref{n1}) and (\ref{n2}), provided that formulae
(\ref{pxi}--\ref{eqxi2}) are taken into
account.

\section{Multicomponent VdW Gas}

The above considerations can be generalized to the multi-component
VdW gas with an arbitrary number of particle species.
After integration over particle momenta and simplifications similar
to those of the above consideration one gets the expression for 
the CPF of
$K$-component VdW gas:
\begin{equation}\label{cpfaK}
Z(V,T, N_1,...,N_K)
~\cong~
\prod_{q=1}^{K}
\frac{1}{N_{q}!}
\left[ \phi (T;m_{q}) \right ]^{N_{q}}
\left( V - \sum_{p=1}^{K} N_{p} \tilde{b}_{pq} \right)^{N_{q}}~,
\end{equation}
where
\begin{eqnarray}
\tilde{b}_{pq}~ &=&~ \frac{2 b_{pp} b_{pq}}{b_{pp}+b_{qq}}~,\\
b_{pq}~& =&~ b_{qp} ~=~ \frac{2}{3} \pi (R_p + R_q)^{3}~.
\end{eqnarray}
$R_q$ is the radius of particle of species$q$. 
$\phi (T;m_{q})$ is defined by Eq.~(\ref{phi}).
The CPF (\ref{cpfaK}) yields the VDW equation of state for
a $K$-component rigid ball gas
\begin{equation}
p(V,T, N_{1},...,N_K )~ =~
\sum_{q=1}^{K}
~\frac{N_{q}T}{V - \sum_{p=1}^{K}  \tilde{b}_{pq}N_{p} }.
\label{steqK}
\end{equation}

The pressure in the grand canonical ensemble is given by the formula
\begin{equation}\label{gcpK}
p(T, \mu_{1},...,\mu_K )~ =~ T~ \sum_{p=1}^{K} \xi_{p}~,
\end{equation}
where $\mu_p$ are the chemical potentials ($p=1,...,K$).
The functions
$\xi_{q}$ satisfy the set of coupled transcendental equations
\begin{equation}
\xi_{p}~ =~ A_{p}~ \exp
\left(
- \sum_{q=1}^{K} \tilde{b}_{pq} \xi_{q}
\right)~,
\end{equation}
where $A_p=\exp(\mu_p/T)\phi(T;m_p)$.
The  particle densities $n_{p}=n_p(T,\mu_1,...,\mu_K)$ are obtained as
the solutions of the following set of coupled linear equations
\begin{equation}
\xi_{p}~ =~ \frac{n_{p}}{1 - \sum_{q=1}^{K} n_{q} \tilde{b}_{qp} }~.
\label{eqnK}
\end{equation}

In the canonical ensemble formulation of the HG model
the numbers $N_1,...,N_K$ are not fixed. They can not be fixed
because of inelastic reactions between the hadrons. The fixed
values have the conserved charges: baryonic number $B$,
strangeness $S$ (strangeness is conserved
as we neglect weak decays) and electric charge $Q$.
The CPF has then the form
\begin{eqnarray}\label{canonical}
Z(V,T,B,S,Q)~&\equiv&~\sum_{N_1,...,N_K=1}^{\infty}
~Z(V,T,N_1,...,N_K)~\delta (B-\sum_{i=1}^{K}b_i N_i)~\\
&\times&~\delta (S-\sum_{i=1}^{K} s_i N_i)~
\delta(Q-\sum_{i=1}^{K} q_iN_i)~,\nonumber
\end{eqnarray}
where $b_i,s_i$ and $q_i$ are the baryonic number, strangeness and
electric charge of $i$-th hadron species.
In the application of the thermal HG models
to A+A collisions, $B$ equals the number of
nucleons participating in the reaction,
$S=0$ and $Q=\alpha B$ with $\alpha \approx 0.5$
for intermediate nuclei and $\alpha \approx 0.4$
for heavy nuclei.
CPF (\ref{canonical}) can be calculated with
the VdW input (\ref{cpfaK}) for $Z(V,T,N_1,...,N_K)$.
The presence of the Kronecker
$\delta$-functions in Eq.~(\ref{canonical}) makes the canonical
ensemble rather complicated. The grand canonical formulation
justified for large systems is much more convenient.
In this case the system properties are defined by the pressure
function (\ref{gcpK}) with the chemical potentials $\mu_i$ ($i=1,...,K$)
defined as
\begin{equation}\label{mui}
\mu_i~=~b_i\mu_B~+~s_i\mu_S~+~q_i\mu_Q~
\end{equation}
in terms of the baryonic $\mu_B$~, strange $\mu_S$~
and electric $\mu_Q$~ chemical potentials.
They are chosen, at given $V$ and $T$, to fix
the {\it average} values $\overline{B}$,
$\overline{S}=0$ and $\overline{Q}=\alpha \overline{B}$.

\section{Summary}

In the present paper we have proposed a generalization
of the VdW excluded
volume procedure for the multicomponent hadron gas.
The canonical and grand canonical ensemble formulations
are presented. Different
hard-core radii for different hadron species
have been discussed in the literature (e.g. [8-12]).
The excluded volume procedure was based on the
substitution $V\rightarrow V-\sum_i b_{ii}N_i$.
This ansatz is shown to be not correct:
for unequal hard-core radii 
the VdW excluded volumes seen by the different hadron species
are different. Thermodynamical quantities
in the grand canonical formulations are found
to be the solution of a coupled set of the transcendental
equations.

\vspace{1cm}

{ \bf Acknowledgements}

We thank K.A. Bugaev, A.P. Kobushkin and  G.~Yen for 
discussion and comments. M.I.G. acknowledges the receipt
of a DFG Professorship at the ITP, Goethe University of
Frankfurt.

\end{document}